\documentclass[prl,aps,twocolumn]{revtex4}

\usepackage{bm}
\usepackage{amsmath}
\usepackage{amssymb}

\usepackage{graphicx}
\begin{document}



\title{Realization of Berezinskii's superconductivity 
in quasi-one-dimensional systems}


\author{Keisuke Shigeta$^{1}$}
\author{Yukio Tanaka$^{1}$}
\author{Kazuhiko Kuroki$^{2}$}
\author{Seiichiro Onari$^{1}$}
\author{Hirohito Aizawa$^{3}$}

\affiliation{$$$^{1}$Department of Applied Physics, 
Nagoya University, Nagoya 464-8603, Japan\\}
\affiliation{$$$^{2}$Department of Applied Physics and Chemistry, 
The University of Electro-Communications, Chofu, Tokyo 182-8585, Japan\\}
\affiliation{$$$^{3}$Institute of Physics, 
Kanagawa University, Rokkakubashi, Yokohama 221-8686, Japan}

\begin{abstract}
We revisit the pairing symmetry competition 
in quasi-one-dimensional systems.
We show that  spin-triplet 
$s$-wave pairing, where the pair is formed by electrons with different times 
and has an odd-frequency symmetry, 
can be realized in systems with strong one-dimensionality 
when the strength of charge fluctuation dominates over spin 
fluctuation. 
The present study provides a novel microscopic mechanism for 
this exotic pairing originally proposed by Berezinskii in 1974. 

\end{abstract}

\pacs{74.20.Mn, 74.20.Rp}

\maketitle
%
%
Superconductivity in strongly correlated systems has been a
long standing issue in condensed matter physics. 
To avoid the strong on-site Coulomb repulsion, 
a Cooper pair tends to be formed by two electrons located in 
separate places with the non-zero angular momentum $L$. 
Spin-singlet $d$-wave ($L=2$) \cite{Miyake}
and spin-triplet $p$-wave ($L=1$) \cite{Schrieffer} pairings
belong to this class. 
On the other hand, two electrons located in separate times 
are also able to avoid the Coulomb repulsion 
in forming a Cooper pair. 
There is a possibility that a sign of the pairing function changes 
in exchanging times of two electrons. 
Then the Fourier transformed pairing function is an odd function 
of frequency.
This type of pairing is called odd-frequency 
(odd-$\mathrm{\omega}$) one originally proposed by Berezinskii 
\cite{Berezinskii}, 
while conventional pairing is called even-frequency 
(even-$\mathrm{\omega}$) one. \par
Berezinskii proposed odd-$\mathrm{\omega}$ 
spin-triplet $s$-wave pairing (Berezinskii's pairing) in a bulk system. 
However, realization of such a pairing has been believed to be
difficult in bulk systems due to the presence of gapless excitation. 
Bergeret $et$ $al.$ revived this exotic pairing in the 
context of ferromagnet / superconductor (F/S) issues \cite{Bergeret}.
They have proposed that Berezinskii's pairing can be  
induced locally 
in ferromagnet although the bulk pairing symmetry in superconductor 
is spin-singlet $s$-wave.  
Stimulated by this proposal there have been several  works 
about this pairing in 
F/S junctions \cite{Ferro1,Ferro2}. 
Besides ferromagnet junctions, 
Berezinskii's pairing has been predicated 
in diffusive normal metal attached to spin-triplet $p$-wave 
superconductor \cite{odd-triplet}. 
It is a really interesting issue to resolve whether 
Berezinskii's pairing is possible in bulk superconductor or not. 
\par
Strictly speaking, there are two classes of odd-$\mathrm{\omega}$ 
superconductors, $i.e.$, spin-singlet and spin-triplet ones. 
Odd-$\mathrm{\omega}$ spin-singlet superconductor \cite{odd1} 
has an odd-parity 
in accordance with Fermi-Dirac statistics. Thus the resulting 
superconducting state is fragile against impurity scattering 
similar to spin-singlet 
$d$-wave and spin-triplet $p$-wave pairing ones. 
Furthermore, temperature dependence of 
Knight shift below transition temperature $T_{\mathrm{C}}$ becomes 
similar to that of spin-singlet $d$-wave superconductor. 
On the other hand, odd-$\mathrm{\omega}$ spin-triplet $s$-wave 
superconductor, the original proposal by Berezinskii, 
can have a clear difference from preexisting pairings, 
since it has an 
unchanged Knight shift below $T_{\mathrm{C}}$ even in the presence of 
impurity scattering. 
Thus, in order to discover odd-$\mathrm{\omega}$ superconductors 
experimentally, 
study on the mechanism of 
Berezinskii's pairing is highly desired. 
Although there have been several studies on the 
generation of odd-$\mathrm{\omega}$ pairing 
up to now \cite{odd1,odd2,odd4}, 
there has not been clear microscopic mechanism 
which supports the realization of Berezinskii's pairing. 
The aim of the present paper is to present a clear microscopic 
mechanism for realizing this exotic pairing.
\par
To resolve this issue, we focus on 
quasi-one-dimensional (Q1D) systems. 
Up to now, superconductivity in Q1D systems
has been studied in the context of, $e.g.$,
an organic superconductor (TMTSF)$_{2}$X.
It has been shown that spin-triplet $f$-wave pairing ($L=3$)
can dominate over spin-singlet $d$-wave ($L=2$) 
one \cite{comment} 
when charge fluctuation dominates over spin fluctuation 
\cite{q1d-1,q1d-2,q1d-3} 
as we shall discuss below. In real space, these correspond to Cooper pairs 
formed between separate places.
However, when the system becomes strongly Q1D, 
it is difficult to form Cooper pairs with separate places 
due to the geometrical constraint, so actually there is a chance that 
even- and odd-$\mathrm{\omega}$ pairings compete. 
In the present paper, combining this ``odd-$\mathrm{\omega}$ $>$ even-$\mathrm{\omega}$'' situation with 
the ``triplet $>$ singlet'' effect, we show that 
spin-triplet $s$-wave ($L=0$) pairing, 
namely, the original Berezinskii's pairing, 
can be realized in a strongly Q1D system with strong charge fluctuation. 
This is actually exemplified by 
solving the linearized Eliashberg's equation in the 
Q1D extended Hubbard model. The present study provides a novel 
and realistic mechanism for realizing this 
exotic pairing proposed more than thirty 
years ago. 
\par
Before going into the actual model and 
the calculation results, we make a general 
argument for the pairing symmetry in a Q1D system. 
We assume a many body system on 
a Q1D lattice, 
where the hopping integral in the $y$ direction $t_y$ is smaller 
than that in the $x$ direction $t_x$. 
The on-site Coulomb repulsion enhances spin fluctuation at the 
nesting vector $\boldsymbol{Q}$. 
In addition, we assume a situation where the off-site 
Coulomb repulsion enhances charge fluctuation at $\boldsymbol{Q}$. 
When the pairing interaction is mainly mediated by 
spin and charge fluctuations at $\boldsymbol{Q}$, 
the effective pairing interactions for spin-singlet and spin-triplet channels 
can be given by
\begin{align}
V_{\mathrm{eff}}^{\mathrm{s}}({\mathrm{i}}\nu_m,\boldsymbol{Q})&=\frac{3}{2}V_{\mathrm{sp}}({\mathrm{i}}\nu_m,\boldsymbol{Q}) 
-\frac{1}{2}V_{\mathrm{ch}}({\mathrm{i}}\nu_m,\boldsymbol{Q}) \label{pairing0s} \\
V_{\mathrm{eff}}^{\mathrm{t}}({\mathrm{i}}\nu_m,\boldsymbol{Q})&=-\frac{1}{2}V_{\mathrm{sp}}({\mathrm{i}}\nu_m,\boldsymbol{Q}) 
-\frac{1}{2}V_{\mathrm{ch}}({\mathrm{i}}\nu_m,\boldsymbol{Q}), \label{pairing0t}
\end{align}
respectively, where $V_{\mathrm{sp}}$ and $V_{\mathrm{ch}}$ are contributions from spin and charge fluctuations, respectively. 
$\nu_m=2m\pi T$ is the bosonic Matsubara frequency with an integer $m$ at the temperature $T$. 
In strongly correlated systems, the effective pairing interaction at $\nu_m=0$ tends to give a large contribution to pairing. 
When the off-site Coulomb repulsion is absent or small, spin-singlet pairing is favored ($|V_{\mathrm{eff}}^{\mathrm{s}}|>|V_{\mathrm{eff}}^{\mathrm{t}}|$) due to the prominence of spin fluctuation. 
We call this ``SF $>$ CF case''. 
On the other hand, when the off-site Coulomb repulsion is 
so remarkable as to make charge fluctuation 
exceed spin fluctuation, spin-triplet pairing is favored 
($|V_{\mathrm{eff}}^{\mathrm{t}}|>|V_{\mathrm{eff}}^{\mathrm{s}}|$). 
We call this ``CF $>$ SF case''. 
It is also noted that the effective pairing interaction for spin-singlet (spin-triplet) channel has a positive (negative) sign as far as the contribution by charge fluctuation does not become too large. 
Since the superconducting gap function $\Delta$ has to satisfy a condition 
$V^{\mathrm{s,t}}(Q)\Delta(k_{\mathrm{F}})\Delta(k_{\mathrm{F}}+Q)<0$ 
on the Fermi surface,
it is required for spin-singlet (spin-triplet) pairing to satisfy a condition $\Delta(k_{\mathrm{F}})\Delta(k_{\mathrm{F}}+Q)<0$ ($\Delta(k_{\mathrm{F}})\Delta(k_{\mathrm{F}}+Q)>0$) with $k_{\mathrm{F}}=({\mathrm{i}}\omega_n,{\boldsymbol{k}}_{\mathrm{F}})$ and 
$Q=(0,{\boldsymbol{Q}})$ consisting of the momentum on the Fermi surface ${\boldsymbol{k}}_{\mathrm{F}}$ and the fermionic Matsubara frequency $\omega_n=(2n-1)\pi T$. 
In the following, we discuss 
all four classes of pairings 
in accordance with the Fermi-Dirac statistics.  
They are
(i) even-$\mathrm{\omega}$ spin-singlet even-parity (ESE),
(ii) even-$\mathrm{\omega}$ spin-triplet odd-parity (ETO),
(iii) odd-$\mathrm{\omega}$ spin-singlet odd-parity (OSO), and
(iv) odd-$\mathrm{\omega}$ spin-triplet even-parity (OTE) pairings.
Berezinskii's pairing belongs to class (iv). 
We consider four cases by combining strongly (weakly) Q1D lattice and SF $>$ CF (CF $>$ SF) case. 
\par
\begin{figure}[htbp]
\includegraphics[width=0.99\linewidth,keepaspectratio]
                  {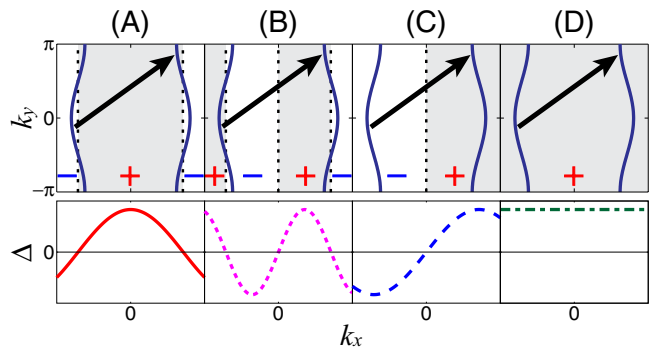}
 \caption{(color online). Pictures of the momentum ${\boldsymbol{k}}=(k_x,k_y)$ dependences of $\Delta$. Upper panels show sign change (positive or negative) of $\Delta$ with nodes (black dashed lines) in momentum space. (Purple) solid curves denote the Fermi surface with the nesting vectors (arrows). Lower panels show the $k_x$ dependences of $\Delta$. (A) $d$-wave in SF $>$ CF case on a weakly Q1D lattice. (B) $f$-wave in CF $>$ SF case on a weakly Q1D lattice. (C) $p$-wave in SF $>$ CF case on a strongly Q1D lattice. (D) $s$-wave in CF $>$ SF case on a strongly Q1D lattice.}
\label{fig_gap}
\end{figure}

First, we discuss a weakly Q1D lattice. 
In this case, electrons avoid the strong Coulomb repulsion in real space. 
Anisotropic pairing with the non-zero angular momentum is induced and 
it has even-$\mathrm{\omega}$ symmetry. 
It is difficult for odd-$\mathrm{\omega}$ pairing to be stabilized due to the nature of pairing with separate times. 
In the SF $>$ CF case, which favors spin-singlet pairing, 
ESE one has an advantage. 
This pairing has $d$-wave symmetry, 
where two nodes of $\Delta$ run close to the Fermi surface as shown in 
Fig. \ref{fig_gap} (A). 
Reversely in the CF $>$ SF case, which favors spin-triplet pairing, 
ETO one dominates. 
This pairing has $f$-wave symmetry, 
where two nodes of $\Delta$ also run close to the Fermi surface as shown in 
Fig. \ref{fig_gap} (B) \cite{triplet2}. 
\par
\begin{figure}[htbp]
  \begin{center}
  \includegraphics[width=0.75\linewidth,keepaspectratio]
                  {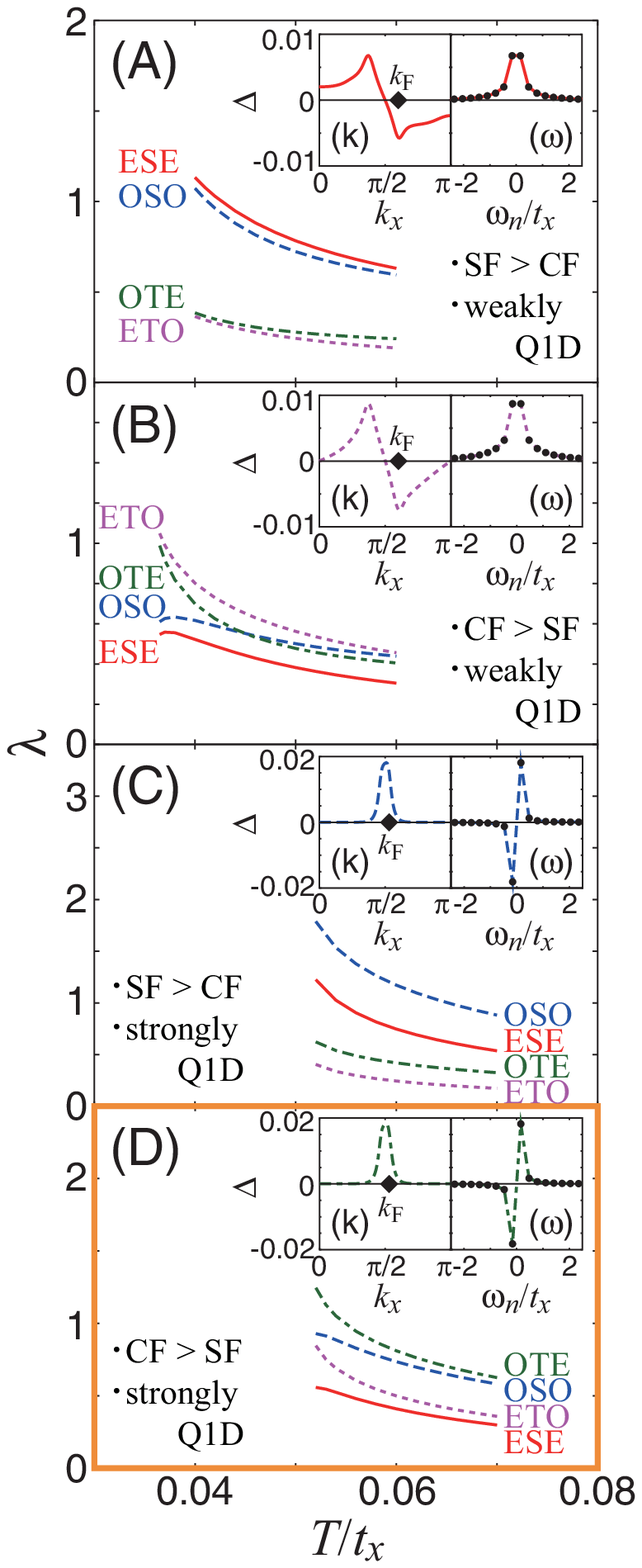}
  \end{center}
  \caption{(color online). $T$ dependences of 
$\lambda$ in SF $>$ CF (CF $>$ SF) case on a weakly (strongly) Q1D lattice. 
Insets (k) and ($\mathrm{\omega}$) are 
${\boldsymbol{k}}=(k_x,k_y)$ and 
$\omega_n$ dependences of $\Delta$ corresponding to the largest eigenvalue, respectively. 
In the inset (k), we choose $k_y=0$ and $\omega_n=\pi T$. 
$k_{\mathrm{F}}$ in the inset denotes the location of the Fermi surface 
at $k_y=0$. 
In the inset ($\mathrm{\omega}$), 
we choose ${\boldsymbol{k}}$ where $\Delta$ gets the largest value on $k_y=0$. }
  \label{fig_T_lambda}
\end{figure}

Next, we consider a strongly Q1D lattice. 
In this case, avoidance of electrons is limited in real space. 
This can induce pairing with separate times and make odd-$\mathrm{\omega}$ 
pairing comparable to even-$\mathrm{\omega}$ one. 
If we focus on the above leading pairings $d$- and $f$-wave, 
the gap nodes will run {\it almost on the entire Fermi surface} (see Fig. \ref{fig_gap} (A),(B)), 
which destabilizes those pairings.
Odd-$\mathrm{\omega}$ symmetry allows pairing 
with lower angular momentum than that of even-$\mathrm{\omega}$ one, 
namely, the number of gap nodes in momentum space is smaller.
As a result of the competition 
between gap nodes in momentum and frequency spaces 
(see the insets of Fig. \ref{fig_T_lambda}), 
odd-$\mathrm{\omega}$ pairing can replace 
even-$\mathrm{\omega}$ one as leading one. 
In the SF $>$ CF case, OSO pairing has an advantage. 
This pairing has $p$-wave symmetry, 
which has no nodes of $\Delta$ on the Fermi surface in momentum space 
as shown in Fig. \ref{fig_gap} (C). 
Finally in the CF $>$ SF case, spin-triplet pairing dominates over spin-singlet one, 
and the OTE pairing can take place. 
This is indeed the spin-triplet $s$-wave pairing originally 
proposed by Berezinskii. 
Here, note that the mechanism is 
entirely novel in that the odd-$\mathrm{\omega}>$ even-$\mathrm{\omega}$ situation due to the strong one-dimensionality 
combined with the triplet $>$ singlet situation 
given by the CF $>$ SF case is the origin of the realization of this exotic 
fully gapped state in momentum space (Fig. \ref{fig_gap} (D)). 
\par
%
%
In order to exemplify the above physics, we actually apply the random phase 
approximation (RPA) to the Q1D extended Hubbard model 
and solve the linearized Eliashberg's equation.
The Hamiltonian considered here is given by 
\begin{align}
{\cal{H}}&=-\sum_{i,j,\sigma}t_{ij}c_{i\sigma}^{\dagger}c_{j\sigma}+\sum_iUn_{i\uparrow}n_{i\downarrow}+\sum_{\langle i,j\rangle}Vn_in_j.
\end{align}
$t_{ij}$ is the hopping integral between sites $i$ and $j$. 
\begin{figure}[htbp]
\includegraphics[width=0.7\linewidth,keepaspectratio]
                  {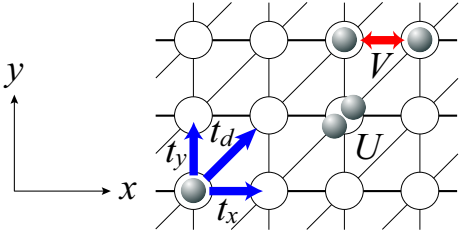}
\caption{(color online). 
The model considered in the present study. $t_x$, $t_y$, and 
$t_d$ are the hopping integrals. 
$U$ and $V$ are the on-site and off-site Coulomb 
repulsions, respectively.}
\label{fig_lattice}
\end{figure}
The hopping integrals between sites neighboring in the $x$, $y$, and diagonal directions are labeled as $t_x$, $t_y$, and $t_d$, respectively as 
shown in Fig. \ref{fig_lattice}. 
We take account of the diagonal hopping integral that 
enhances the geometrical frustration, which can significantly induce 
competition between even- and odd-$\mathrm{\omega}$ pairings. 
The dispersion is given by 
$\varepsilon_{\boldsymbol{k}}=-2t_x\cos k_x-2t_y\cos k_y-2t_d\cos(k_x+k_y)$. 
We choose $t_y=t_d=0.35t_x$ ($t_y=t_d=0.1t_x$) as the hopping integrals for a weakly (strongly) Q1D lattice. 
$c_{i\sigma}^{(\dagger )}$ and $n_{i\sigma}$ are the annihilation (creation) and number operators for an electron with spin $\sigma$ on a site $i$. 
$n_i=n_{i\uparrow}+n_{i\downarrow}$. 
$U$ and $V$ are the on-site and off-site Coulomb repulsions, respectively. 
$V$ acts between electrons neighboring in the $x$ direction. 
The momentum dependence is given by $V({\boldsymbol{q}})=2V\cos q_x$ with the momentum ${\boldsymbol{q}}=(q_x,q_y)$. 
In this model, we solve the linearized Eliashberg's equation for spin-singlet (spin-triplet) channel within the RPA
\begin{eqnarray}
\lambda\Delta(k)=-\frac{T}{N}\sum_{k'}V_{\mathrm{eff}}^{\mathrm{s(t)}}(k-k')|G_0(k')|^2\Delta(k'),
\end{eqnarray}
where $N$ is the number of sites and 
$k\equiv({\mathrm{i}}\omega_n,{\boldsymbol{k}})$. 
$G_0(k)=({\mathrm{i}}\omega_n-\varepsilon_{\boldsymbol{k}}+\mu)^{-1}$ is the bare Green's function with the chemical potential $\mu$. 
$\lambda$ is the eigenvalue for $\Delta$. 
$\lambda$ becomes unity just at $T_{\mathrm{C}}$. 
The more stable the superconducting state is, the larger $\lambda$ tends to be. 
We calculate $\lambda$ and $\Delta$ for ESE, ETO, OSO, and OTE symmetries. 
The effective pairing interactions for spin-singlet and spin-triplet channels within the RPA are given by
\begin{align}
V_{\mathrm{eff}}^{\mathrm{s}}(q)&=U+V({\boldsymbol{q}})+\frac{3}{2}U^2\chi_{\mathrm{sp}}(q) \notag \\
&\mspace{80mu}-\frac{1}{2}\{U+2V({\boldsymbol{q}})\}^2\chi_{\mathrm{ch}}(q) \label{pairing1} \\
V_{\mathrm{eff}}^{\mathrm{t}}(q)&=V({\boldsymbol{q}})-\frac{1}{2}U^2\chi_{\mathrm{sp}}(q) \notag \\
&\mspace{80mu}-\frac{1}{2}\{U+2V({\boldsymbol{q}})\}^2\chi_{\mathrm{ch}}(q), \label{pairing3}
\end{align}
respectively, with $q\equiv({\mathrm{i}}\nu_m,{\boldsymbol{q}})$. 
The spin and charge susceptibilities are given by
\begin{eqnarray}
\chi_{\mathrm{sp}}(q)&=&\frac{\chi_0(q)}{1-U\chi_0(q)} \\
\chi_{\mathrm{ch}}(q)&=&\frac{\chi_0(q)}{1+\{U+2V({\boldsymbol{q}})\}\chi_0(q)},
\end{eqnarray}
respectively, with the irreducible susceptibility
\begin{eqnarray}
\chi_0(q)=-\frac{T}{N}\sum_{k}G_0(q+k)G_0(k).
\end{eqnarray}
In the present paper, we normalize the gap function $\sum_{k}|\Delta(k)|/N=1$. 
We take the number of sites $N=N_x\times N_y=256\times64$. 
The number of electrons is unity per site (half-filled), 
which gives the nesting vector ${\boldsymbol{Q}}=(\pi,\pi/2)$ 
both on weakly and strongly Q1D lattices. 
The fermionic and bosonic Matsubara frequency have values from $-(2N_{\mathrm{f}}-1)\pi T$ to $(2N_{\mathrm{f}}-1)\pi T$ and from $-2N_{\mathrm{f}}\pi T$ to $2N_{\mathrm{f}}\pi T$, respectively, with $N_{\mathrm{f}}=2048$. 
\par
%
%
In Fig. \ref{fig_T_lambda}, $\lambda$ for four kinds of pairings are plotted against $T$ in the four cases. 
There are two insets in panels Fig. \ref{fig_T_lambda}(A)-(D), 
where (k) $k_{x}$ and ($\mathrm{\omega}$) $\omega_{n}$ dependences 
of $\Delta$ are plotted at $T=0.05t_x$. 
In the inset (k), we choose $k_y=0$ and $\omega_n=\pi T$. 
$k_{\mathrm{F}}$ in the inset denotes the location of the Fermi surface 
at $k_y=0$. 
In order to give 
clear comparison with inset ($\mathrm{\omega}$), we plot only the 
$k_x\geq 0$ portion, while the $k_x<0$ portion 
is given by $\Delta(\boldsymbol{k})=+(-)\Delta(-\boldsymbol{k})$ for 
even- (odd-) parity pairings.
In the inset ($\mathrm{\omega}$), we choose ${\boldsymbol{k}}$ where $\Delta$ gets the largest value on $k_y=0$. 
\par
First, we focus on a weakly Q1D lattice ($t_y=t_d=0.35t_x$). 
In the SF $>$ CF case ($U=2t_x$, $V=0$), 
ESE pairing is the most stable 
as shown in Fig. \ref{fig_T_lambda}(A). 
This pairing has $d$-wave gap, whose $k_x\geq 0$ portion is shown in the inset (k) of Fig. \ref{fig_T_lambda}(A). 
There is a node of $\Delta$ near $k_x=k_{\mathrm{F}}$ in momentum space, while there are no nodes in Matsubara frequency space as shown in the inset ($\mathrm{\omega}$) of Fig. \ref{fig_T_lambda}(A). 
In the CF $>$ SF case ($U=1.995t_x$, $V=t_x$), 
ETO pairing is the most stable 
as shown in Fig. \ref{fig_T_lambda}(B). 
This pairing has $f$-wave gap, whose $k_x\geq 0$ portion is shown in the inset (k) of Fig. \ref{fig_T_lambda}(B) \cite{q1d-1}. 
There is also a node of $\Delta$ near $k_x=k_{\mathrm{F}}$ in momentum space, while there are no nodes in Matsubara frequency space as shown in the inset ($\mathrm{\omega}$) of Fig. \ref{fig_T_lambda}(B). 
\par
\begin{figure}[htbp]
 \begin{center}
  \includegraphics[width=0.84\linewidth,keepaspectratio]
                  {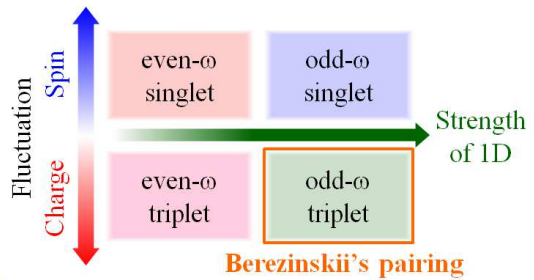}
  \end{center}
  \caption{(color online). 
The most stable pairings dependent on the strength of one-dimensionality 
and spin/charge fluctuation.}
  \label{fig_summary}
\end{figure}

Next, we focus on a strongly Q1D lattice ($t_y=t_d=0.1t_x$). 
In the SF $>$ CF case ($U=1.6t_x$, $V=0$), 
OSO pairing is the most stable 
as shown in Fig. \ref{fig_T_lambda}(C). 
This pairing has $p$-wave gap, whose $k_x\geq 0$ portion is shown in the inset (k) of Fig. \ref{fig_T_lambda}(C) \cite{odd4}. 
There are no nodes of $\Delta$ near $k_x=k_{\mathrm{F}}$ in momentum space, and instead there is a node in Matsubara frequency space as shown in the inset ($\mathrm{\omega}$) of Fig. \ref{fig_T_lambda}(C). 
And finally for the CF $>$ SF case ($U=1.595t_x$, $V=0.8t_x$), 
OTE pairing is indeed the most stable 
as shown in Fig. \ref{fig_T_lambda}(D). 
This pairing has $s$-wave gap, whose $k_x\geq 0$ portion is shown in the inset (k) of Fig. \ref{fig_T_lambda}(D). 
There are also no nodes of $\Delta$ near $k_x=k_{\mathrm{F}}$ in momentum space, and instead there is a node in Matsubara frequency space as shown in the inset ($\mathrm{\omega}$) of Fig. \ref{fig_T_lambda}(D). It is remarkable that 
the realization of 
Berezinskii's pairing has been verified based on a microscopic calculation. 
These four cases are summarized in Fig. \ref{fig_summary}. 
\par
%
%
To summarize, 
we have shown 
that odd-$\mathrm{\omega}$ spin-triplet $s$-wave pairing originally proposed by Berezinskii 
can be realized 
in systems with strong one-dimensionality when 
the strength of charge fluctuation exceeds over 
that of spin fluctuation. 
Experimentally, it is interesting to look for this exotic pairing 
in Q1D materials where spin and charge fluctuations coexist. 
Moreover, we may expect applying magnetic field to enhance 
spin-triplet pairing (even when 
charge fluctuation is not so strong) \cite{q1d-3}, so combining this effect 
with the strong one-dimensionality may increase chances for 
realizing Berezinskii's pairing in actual materials. 
Also, we hope anomalous properties specific to 
this pairing will be observed in Q1D systems \cite{odd1,Linder}. 
\par
%
%
This work was supported by a Grant-in-Aid for 
Scientific Research (Grant No.\ 20654030) from 
MEXT, Japan. K.S. acknowledges support by JSPS.
%
%


\begin{thebibliography}{99}
\bibitem{comment}
Here the ``angular momentum'' $L$ for the Q1D case 
is not used in its strict sense since the nodes of the gap do not 
necessarily pass through the origin. Nonetheless, we use this term 
meaning the number of nodes that intersect or run nearby the 
Fermi surface.
\end{thebibliography}
\end{document}